\documentclass[preprint,onecolumn,nofootinbib]{revtex4-1}
\usepackage[colorlinks=true,linkcolor=blue,urlcolor=blue,filecolor=black,citecolor=red,pdfstartview=FitV,pdftitle={},pdfsubject={},pdfkeywords={},pdfpagemode=None,bookmarksopen=true]{hyperref}
\usepackage{graphicx}
\usepackage{amsmath}
\usepackage{amssymb,ulem}
\usepackage{color}%
\usepackage{tikz}
\usepackage{dcolumn}
\usepackage{bbold}
\usepackage{times}
\usepackage{pdfsync}
\usepackage{xcolor}
\usepackage{setspace}
\usepackage[T1]{fontenc}

\newcommand{\SOUTHCUT}{$^1$ School of Physics and Optoelectronics, South China University of Technology, Guangzhou 510641,People's Republic of China}
\newcommand{\JINAN}{$^2$ Department of Physics and Siyuan Laboratory, Jinan University, Guangzhou 510632, China}
\begin{document}
\title{The rotating solutions beyond the spontaneous scalarization in Einstein-Maxwell-scalar theory}
\author{Wei Xiong $^{1}$}
\email{202210187053@mail.scut.edu.cn}
\author{Cheng-Yong Zhang $^{2}$}
\email{zhangcy@email.jnu.edu.cn}
\author{Peng-Cheng Li $^{1}$}
\email{pchli2021@scut.edu.cn, corresponding author}

\affiliation{\SOUTHCUT}
\affiliation{\JINAN}
\begin{abstract}
    The Einstein-Maxwell-scalar (EMS) theory with a quartic coupling function features three branches of fundamental black hole (BH) solutions, labeled as cold, hot, and bald black holes. The static bald black holes (the Reissner-Nordström BH) exhibit an intriguing nonlinear instability beyond the spontaneous scalarization. We study the rotating scalarized black hole solutions in the EMS model with a quartic coupling function through the spectral method numerically. The domain of existence for the scalarized BHs is presented in the spin-charge region. We found that the rotating solutions for both the two scalarized branches possess similar thermodynamic behavior compared to the static case while varying the electric charge. The BH spin enlarges the thermodynamic differences between the cold and hot branches. The profile of the metric function and the scalar field for the scalarized BHs is depicted, which demonstrates that the scalar field concentrates more on the equatorial plane in contrast to the axisymmetric region as the spin increases.
\end{abstract}
\maketitle
    
\section{Introduction}
\label{section1}

In recent years, the precise measurement of black holes in strong field regions has become achievable through advanced gravitational wave observations by the LIGO/Virgo collaboration\cite{LIGOScientific:2018mvr,LIGOScientific:2020ibl,LIGOScientific:2021djp} and black hole images by the Event Horizon Telescope collaboration\cite{EventHorizonTelescope:2019dse,EventHorizonTelescope:2019ggy,EventHorizonTelescope:2019jan,EventHorizonTelescope:2019pgp,EventHorizonTelescope:2019ths,EventHorizonTelescope:2019uob}. 
General relativity (GR), despite its precise validation in the weak field region, is currently undergoing the most rigorous test ever near the BH horizon\cite{Berti:2015itd,Ghosh:2022xhn,LIGOScientific:2021sio,Niu:2022yhr}. 
Many theorems and assumptions within GR necessitate a further examination (e.g., the symmetries of Lorentz invariance \cite{Mattingly:2005re}). 
In the framework of GR, the well-known no-hair theorem declares that the BH solution as a outcome of gravitational collapse is of the Kerr-Newman (KN) families uniquely\cite{Ruffini:1971bza}. The KN BHs are electrovacuum and prohibit the stationary existence of other matter fields. However, a stationary BH with a nontrivial scalar configuration can be generated while relaxing the restriction of the assumptions within the no-hair theorem through a physical motivation. One family of those models that has received much interest is spontaneous scalarization, as it  vanishes in the weak field and emerges in the strong field region\cite{Doneva:2022ewd}. That explains the reason that we have not yet detected this phenomenon and highlights the potential application of spontaneous scalarization for future advanced detection in the strong field region. 

The concept of spontaneous scalarization was first introduced in the Einstein-scalar-Gauss-Bonnet (EsGB) model, which involves a nonminimal coupling between the scalar field and the Gauss-Bonnet curvature term\cite{Sotiriou:2013qea,Antoniou:2017acq,Doneva:2017bvd,Antoniou:2017hxj,Doneva:2018rou,Belkhadria:2023ooc}. This nonminimal coupling imparts an effective mass square term to the scalar field. Given certain parameters for this model, the effective mass square is sufficiently negative to trigger the tachyonic instability of the scalar field on the background of the BH\cite{Blazquez-Salcedo:2018jnn}. The Schwarzschild BH in EsGB model is hence unstable and spontaneously transitions into a stable scalarized BH solution after undergoing an arbitrarily small perturbation. Besides being induced by a curvature source, the spontaneous scalarization can also be triggered by a matter source such as the electromagnetic field\cite{Herdeiro:2018wub,Fernandes:2019rez}. Such a model is referred to as the Einstein-Maxwell-scalar (EMS) model, which brings about the spontaneous scalarization in a more simple manner. In EMS model, the static scalarized BHs are first constructed using the shooting method with the spherical symmetry, and subsequently regained through the fully nonlinear numerical evolution\cite{Zhang:2021ybj,Xiong:2022ozw,Niu:2022zlf}. After giving an initial perturbation to a Reissner-Nordström (RN) black hole in EMS models, the scalar field exponentially grows and eventually forms a static configuration.

A new mechanism beyond spontaneous scalarization can also lead to the scalarized BHs through a nonminimal coupling between scalar and a source term\cite{Doneva:2021tvn,Lai:2023gwe,Blazquez-Salcedo:2020nhs,Zhang:2021nnn,Liu:2022fxy,Jiang:2023yyn,Promsiri:2023yda}. It features a coupled function that exceeds quadratic for the scalar, whereas the coupled function related to the spontaneous scalarization is quadratic in the scalar field. Then the scalar perturbation on the background of vacuum BH avoids the tachyonic instability, because the scalar perturbation equation returns to the form in GR. The scalar field of the vacuum BH (Schwarzschild BH) must undergo sufficiently large scalar perturbation to trigger a nonlinear instability, which was first reported in the EsGB model beyond spontaneous scalarization\cite{Doneva:2021tvn}. However, this study is restricted to the decoupling limit which means that the background of Schwarzschild BH is fixed under evolution. The full backreaction numerical evolution was first achieved in the EMS model beyond spontaneous scalarization\cite{Zhang:2021nnn}. There are three families of fundamental solutions, including two stable branches (hot BHs and bald BHs) and one unstable branch (cold BHs) under perturbations\cite{LuisBlazquez-Salcedo:2020rqp}. The bald BH solutions are the RN BHs without scalar hair and BH solutions of other branches are hairy.

The existence of BH spin has been strongly supported by gravitational wave observations\cite{LIGOScientific:2018mvr,LIGOScientific:2020ibl,LIGOScientific:2021djp} and BH images currently\cite{Cui:2023uyb}. It is more realistic to incorporate angular momentum into the research of scalarized BHs than working in the static limit. A comprehensive investigation of the effects of BH spin can assist us in better constraining coupled parameters of models mentioned above, using astronomical observational data\cite{Silva:2022srr,Wang:2021jfc}. In general, the BH spin restricts the parameter range of rotating hairy BHs for the spontaneous scalarization coupling where the scalar condensation is induced by the source term\cite{Kleihaus:2011tg,Cunha:2019dwb,Guo:2023mda}. In the spin-induced spontaneous scalarization, BH spin expands the parameter range for hairy solutions that do not possess a static limit\cite{Dima:2020yac,Herdeiro:2020wei,Berti:2020kgk}. The nonlinear scalarized rotating BHs in EsGB model beyond spontaneous scalarization are presented in \cite{Lai:2023gwe}. The effect of BH spin on the EMS model beyond spontaneous scalarization is still unknown. The thermodynamic investigation of scalarized BHs beyond spontaneous scalarization can provide revelation for the dynamic properties of the nonlinear scalarization while considering the numerical evolution under the axisymmetric spacetime.

In this paper, we study the rotating scalarized BHs, including cold and hot branches, in the EMS model beyond spontaneous scalarization. The parameter region for both cold and hot branches on the spin-charge plane is demonstrated. The thermodynamic behaviors with varying parameters are investigated and the function profiles for the scalarized BH are presented. We also show the error estimation established by the Smarr relation for both cold, hot, and bald BHs. This paper is organized as follows. In section.\ref{section2}, we introduce the EMS model in stationary axisymmetric spacetime and the spectral method utilized to calculate numerical hairy solutions. The results are carried out in section.\ref{section3}. We discuss in section.\ref{section4}. By convention, the geometric units $G=c=1$ are employed. 

\section{Background}
\label{section2}

\subsection{Einstein-Maxwell-scalar model}

The following action decribes the Einstein-Maxwell-scalar model\cite{Fernandes:2019rez}
\begin{equation}
    S = \frac{1}{4\pi} \int d^{4}x \sqrt{-g} \left[R- 2 \nabla_{\mu}\phi\nabla^{\mu}\phi- f(\phi) F_{\mu\nu}F^{\mu\nu} \right],
	\label{eq:action}
\end{equation}
where $R$ is the Ricci scalar, $\phi$ is a real scalar field and $F_{\mu\nu}$ is the Maxwell 2-form. A nonminimal coupling exists between the coupling function $f(\phi)$ and the Maxwell invariant $F_{\mu\nu}F^{\mu\nu}$. In the context of spontaneous scalarization mechanism, coupling functions need to satisfy corresponding constraints
\begin{equation}
    f(0) = 1, \  \frac{df}{d\phi}(0) = 0, \ \frac{d^{2}f}{d\phi^{2}}(0) >0. 
\end{equation}
A typical coupling function for the spontaneous scalarization has the Taylor expansion $1-b\phi^{2}+\ldots$ around $\phi=0$, for example, $f(\phi)=e^{-b\phi^{2}}$, where $b$ is the coupling constant. To satisfy the above constraints, $b$ must be chosen as a negative number. However, Hod $et \ al.$ discovered a spin-induced spontaneous scalarization phenomenon in EMS models while selecting $b$ as a positive number\cite{Hod:2022txa}. Lai $et \ al.$  also verified this phenomenon through the linear evolution of the perturbed scalar field on the background of KN BHs in this model\cite{Lai:2022spn,Lai:2022ppn}. The tachyonic instability of the scalar field is triggered by a nonzero BH spin exceeding a certain threshold. Otherwise, a different coupling function beyond the spontaneous scalarization can be expanded as the form $1-b\phi^{4}+\ldots$ around 0, e.g., $f(\phi)=e^{-b\phi^{4}}$\cite{Zhang:2021nnn,Blazquez-Salcedo:2020nhs}. Such type of coupling function obeys the constraint
\begin{equation}
    f(0) = 1, \  \frac{df}{d\phi}(0) = 0, \ \frac{d^{2}f}{d\phi^{2}}(0) =0. 
\end{equation}
The quartic subleading term of $\phi$ leads to the existence of a branch of stable bald black hole solutions (KN BHs). The differences between the spontaneous scalarization (the quadratic coupling) and the beyond spontaneous scalarization (the quartic coupling) can be revealed by the scalar equation of motion
\begin{equation}
	\nabla^{\mu}\nabla_{\mu}\phi-\frac{1}{4} \frac{d f(\phi)}{d \phi} F^{\mu\nu}F_{\mu\nu}= 0. 
	\label{eq:scalar}
\end{equation}
The scalar field becomes the perturbation $\phi=\delta \phi$ on the background of bald BHs. For the coupling function $f(\phi)=e^{-b\phi^{2}}$ of the spontaneous scalarization, the derivative $\frac{d f(\phi)}{d \phi} = -2b\phi \  e^{-b\phi^{2}}$ gives $-2b\delta\phi$ while considering the linear approximation. The second term in (\ref{eq:scalar}) becomes a negative effective mass term within certain parameter region, which induces a tachyonic instability of the scalar perturbation destabilizing the bald BHs. However, the derivative $\frac{d f(\phi)}{d \phi}$ vanishes with respect to the coupling function $f(\phi)=e^{-b\phi^{4}}$ in the linear level. The scalar perturbation of Kerr-Newman black holes returns to that of the Einstein-Maxwell model, known as linearly stable.

We use a specific ansatz for the metric followed by\cite{Herdeiro:2014goa,Herdeiro:2015gia,Guo:2023mda}
\begin{eqnarray}
	ds^{2}&= &-e^{2 F_{0}(r,\theta)} (1-r_{h}/r) dt^{2} + e^{2F_{1}(r,\theta)} \left( \frac{dr^{2}}{(1-r_{h}/r)}+r^{2} d\theta^{2} \right) \nonumber \\
        & & + e^{2F_{2}(r,\theta)} r^{2} \sin^{2}\theta (d\varphi-\frac{W(r,\theta)}{r^{2}}dt)^{2} \label{eq:ansatz1} \\
    A_{\mu} dx^{\mu} &=& \left(A_{t}(r,\theta)-\frac{W(r,\theta)}{r^{2}}A_{\varphi}(r,\theta) \sin^{2}\theta \right) dt +A_{\varphi}(r,\theta) \sin^{2}\theta \ d\varphi,
    \label{eq:ansatz2}
\end{eqnarray}
with two Killing vectors $\xi=\partial_{t}$ and $\eta=\partial_{\varphi}$. This ansatz has been used for the rotating boson stars with the event horizon radius $r_{h}=0$\cite{Schunck:1996he,Yoshida:1997qf} and describes a stationary, axisymmetric, and asymptotically flat BH while $r_{h} \neq 0$. So the radial domain is given by $r_{h} \leq r \leq \infty$. We focus on the solutions which are symmetric with respect to a reflection on the equatorial plane $\theta = \pi/2$ and hence can restrict the $\theta$ in $[0,\pi/2]$. The scalar field $\phi(r,\theta)$ inherits the spacetime symmetries. The transformation from the Boyer-Lindquist coordinates to this coordinates system is given By
\begin{equation}
    r_{\textrm{BL}} = r+M-\sqrt{M^{2}-Q^{2}-a^{2}},
    \label{eq:coordinate}
\end{equation}
where $r_{\textrm{BL}}$ denotes the radial coordinate of the Boyer-Lindquist system. 

We denote the Einstein equation and the Maxwell equation as
\begin{eqnarray}
    E^{\mu}_{\nu} &\equiv& G^{\mu}{}_{\nu}- T^{\mu}{}_{\nu} =0 , \label{eq:Einstein}\\
    M_{\nu} &\equiv&  \nabla^{\mu}[f(\phi) F_{\mu\nu}]  =0 , \label{eq:Maxwell}
\end{eqnarray}
with the energy-momentum tensor 
\begin{equation}
    T^{\mu}{}_{\nu}  \equiv 2 \left[ \nabla^{\mu}\phi\nabla_{\nu}\phi -\frac{1}{2} \delta^{\mu}{}_{\nu} \nabla^{\rho}\phi\nabla_{\rho}\phi +  f(\phi) (F^{\mu\rho}F_{\nu\rho}-\frac{1}{4}\delta^{\mu}{}_{\nu} F^{\sigma\rho}F_{\sigma\rho}) \right].
    \label{eq:energy momentum tensor}
\end{equation}
Above equations can be recombined as the following form
\begin{eqnarray}
    E^{r}_{r}+E^{\theta}_{\theta}+E^{\varphi}_{\varphi}-E^{t}_{t}-\frac{2W}{r^{2}}E^{t}_{\varphi} &=& 0, \nonumber \\
    E^{r}_{r}+E^{\theta}_{\theta}-E^{\varphi}_{\varphi}+E^{t}_{t}+\frac{2W}{r^{2}}E^{t}_{\varphi} &=& 0, \nonumber \\
    E^{r}_{r}+E^{\theta}_{\theta}-E^{\varphi}_{\varphi}-E^{t}_{t}                                 &=& 0, \nonumber \\
    E^{t}_{\varphi}                                                                               &=& 0,  \nonumber \\
    M_{t}+\frac{W}{r^{2}}M_{\varphi} - \frac{2(r-r_{h})A_{\varphi}}{r^{3}} e^{2F_{0}-2F_{2}} f(\phi) E^{t}_{\varphi} &=& 0, \nonumber\\
    M_{\varphi}                                                                                   &=& 0.
	\label{eq:equations}                                                                                  
\end{eqnarray}
These equations together with the Klein-Gordon equation (\ref{eq:scalar}) describe a set of seven nonlinear, coupled, second-order PDEs with seven functions ($F_{0},F_{1},F_{2},W,\phi,A_{t},A_{\varphi}$). Each of these equations have the second derivatives $\partial_{r}\partial_{r}F(r,\theta)+\frac{\partial_{\theta}\partial_{\theta}F(r,\theta)}{r^{2}-r\ r_{h}}$ of a single function, where $F(r,\theta)$ represents one of the seven unknown functions. 

The asymptotic behaviors of functions at infinity 
\begin{equation}
    e^{2F_{0}}(1-\frac{r_{h}}{r}) \sim 1-\frac{2M}{r} ,\ \ W \sim \frac{2J}{r}, \ \ \phi \sim \frac{Q_{s}}{r} , \ \ A_{t} \sim \Phi-\frac{Q}{r} , \ \ \textrm{while}\ r\rightarrow\infty
    \label{eq:asymptotic}
\end{equation}
manifest three conserved charges (BH mass $M$, BH angular momentum $J$ and electric charge $Q$) related to the symmetries of fields. The scalar charge $Q_{s}$ has been found to rely on other conserved charges\cite{Fernandes:2019rez}. This implies that the hairy BHs in EMS model are secondary\cite{Herdeiro:2015waa}. $\Phi$ is the electrostatic potential selected to impose $A_{t}=0$ at the horizon. We define the dimensionless spin $\chi \equiv J/M^{2}$ and the dimensionless charge $q \equiv Q/M$ by the BH mass. The Hawking temperature and the horizon area is given by\cite{Herdeiro:2014goa}
\begin{equation}
    T_{h}=\frac{1}{4\pi r_{h}} e^{F_{0}(r_{h},\theta)-F_{1}(r_{h},\theta)}, \ \  A_{h} = 2 \pi r_{h}^{2} \int_{0}^{\pi} d\theta \sin \theta e^{F_{1}(r_{h},\theta)+F_{2}(r_{h},\theta)},
\end{equation}
respectively. We also introduce the dimensionless version of quantities above: the reduced temperature $t_{h} \equiv 8\pi M T_{h}$ and the reduced horizon area $a_{h}\equiv \frac{A_{h}}{16 \pi M^{2}}$. The Smarr relation\cite{Townsend:1997ku} is
\begin{equation}
    M=\Phi Q + \frac{1}{2} T_{h} A_{h}+2 \Omega_{h} J,
    \label{eq:Smarr}
\end{equation}
where $\Omega_{h}$ denotes the horizon angular momentum. The deviation of Smarr relation (\ref{eq:Smarr}) for numerical solutions is utilized to assess the numerical accuracy of our code.

\subsection{The method}
We redefine the radial coordinate as
\begin{equation}
    x\equiv\frac{\sqrt{r^{2}-r_{h}^{2}}-r_{h}}{\sqrt{r^{2}-r_{h}^{2}}+r_{h}},
    \label{eq:redefinition of r}
\end{equation}
to project the domain of radial $r \in [r_{h},\infty)$ to $x \in [-1,1]$, where $x=-1$ and $x=1$ corresponding to the event horizon and the infinity\cite{Guo:2023mda}. The compactification of the coordinate domain is to accommodate the spectral method mentioned below. Then we can impose the following boundary conditions for the radial part
\begin{eqnarray}
    \partial_{x}F_{0}=\partial_{x}F_{1}=\partial_{x}F_{2}=\partial_{x}W=\partial_{x}\phi=A_{t}=\partial_{x}A_{\varphi}=0 &,\ x=-1 \label{eq:boundary radial1} \\
    F_{0}=F_{1}=F_{2}=\partial_{x}W+\frac{r_{h}}{4}(1+4\partial_{x}F_{0})^{2}\chi=\phi=& \nonumber \\
    4\partial_{x}A_{t}-q(1+4\partial_{x}F_{0})=A_{\varphi}=0 &,\ x=1.
    \label{eq:boundary radial2}
\end{eqnarray}
by considering the asymptotic behaviors (\ref{eq:asymptotic}). 
We select the input pair $(r_h, b, \chi, q)$, which corresponds to four adjustable parameters for this system. The dimensionless spin $\chi$ and electric charge $q$ comes from the monopole moment of the asymptotic expansion (\ref{eq:asymptotic}) for function $W$ and $A_{t}$ at infinity respectively. There are also other quantities (for example, $\Omega_{h}$ and $\Phi$ from the expansion for $W$ and $A_{t}$ at event horizon) as the adjustable parameters. The boundary conditions for the angular part can be fixed by the ansatz (\ref{eq:ansatz1}) and (\ref{eq:ansatz2}), 
\begin{equation}
    \partial_{\theta}F_{0}=\partial_{\theta}F_{1}=\partial_{\theta}F_{2}=\partial_{\theta}W=\partial_{\theta}\phi=\partial_{\theta}A_{t}=\partial_{\theta}A_{\varphi}=0, \ \ \ \theta =0 ,  \frac{\pi}{2}.
    \label{eq:boundary angular}
\end{equation}

The seven nonlinear equations (\ref{eq:scalar}) and (\ref{eq:equations}) together with the physical boundary conditions (\ref{eq:boundary radial1}), (\ref{eq:boundary radial2}) and (\ref{eq:boundary angular}) are solved by the spectral method in this paper. We implement the numerical approach introduced in \cite{Dias:2015nua,Fernandes:2022gde,Garcia:2023ntf}. The functions of the nonlinear system are decomposed into the spectral expansion with the Chebyshev polynomials and the trigonometric functions in the spectral method, 
\begin{equation}
    F(x,\theta) = \sum^{N_{x}-1}_{n=0} \sum^{N_{\theta}-1}_{m=0} \alpha_{nm} \ T_{n}(x) Y_{m}(\theta),
    \label{eq:expansion}
\end{equation}
where $F$ represents any function ($F_{0},F_{1},F_{2},W,\phi,A_{t},A_{\varphi}$) in this system to be soloved. $N_{x}$ and $N_{\theta}$ denote the resolution of $x$ and $\theta$ respectively. $\alpha_{nm}$ is the expansion coefficient. The $n$th Chebyshev polynomial is defined as $T_{n}(x) \equiv \cos (n \Theta)$ with $\Theta \equiv \arccos x$. The Chebyshev polynomials are guaranteed to converge exponentially fast while approximating a aperiodic smooth function defined in the interval $x \in [-1,1]$\cite{Fernandes:2022gde}. 
$Y_{m}(\theta)$ can be choosen to the trigonometric functions $\cos (2m \theta)$ uniformly and hence the angular boundary conditions (\ref{eq:boundary angular}) are satisfied automatically. The functions are approximated by the interpolation composed of series (\ref{eq:expansion}) and the collocation points, which are defined as
\begin{eqnarray}
    x_{i}=& \cos \left(\frac{\pi}{2} \frac{2i+1}{N_{x}-2}\right) &,\ \  i=0, \ldots ,N_{x}-3,   \\
    \theta_{j}=& \frac{(2j+1) \pi}{4N_{\theta}} &, \ \ j=0,\ldots,N_{\theta}-1.
    \label{eq:points}
\end{eqnarray}

Utilizing the approximation (\ref{eq:expansion}) and the discretization (\ref{eq:points}), one can convert the nonlinear PDE problem to the root finding problem of the nonlinear coupled algebraic equations with variables $\alpha_{nm}$, which is called the spectral method. Remind that the angular boundary conditions (\ref{eq:boundary angular}) are satisfied automatically, so (\ref{eq:boundary angular}) is no need to impose in the calculation. However, it is much beneficial for the convergence of the code while imposing the supplemental condition $F_{1}=F_{2}$ for the removal of the conical singularity. Such a complex system can not be solved in an analytic process. The Newton-Raphson method is employed to iteratively approach the root with a suitable initial value $\alpha_{nm}^{(0)}$. The choice of the initial value plays a crucial role in ensuring the convergence of the Newton-Raphson method. Fortunately, the solutions of the quartic coupling function possess a static limit, allowing us to use the shooting method to obtain a static solution just as in the previous works\cite{Blazquez-Salcedo:2020nhs}. The build-in program $LinearSolve$ in Mathematica is utilized to solve the matrix equation resulting from the Newton-Raphson method. We can then adjust parameters slightly to traverse the entire existence domain step by step. We fix the resolution $(N_{x},N_{\theta})$ = (42,8) by convention. 

\section{Results}
\label{section3}

\begin{figure}[htbp]
    \centering
    \includegraphics[width = 0.5\textwidth]{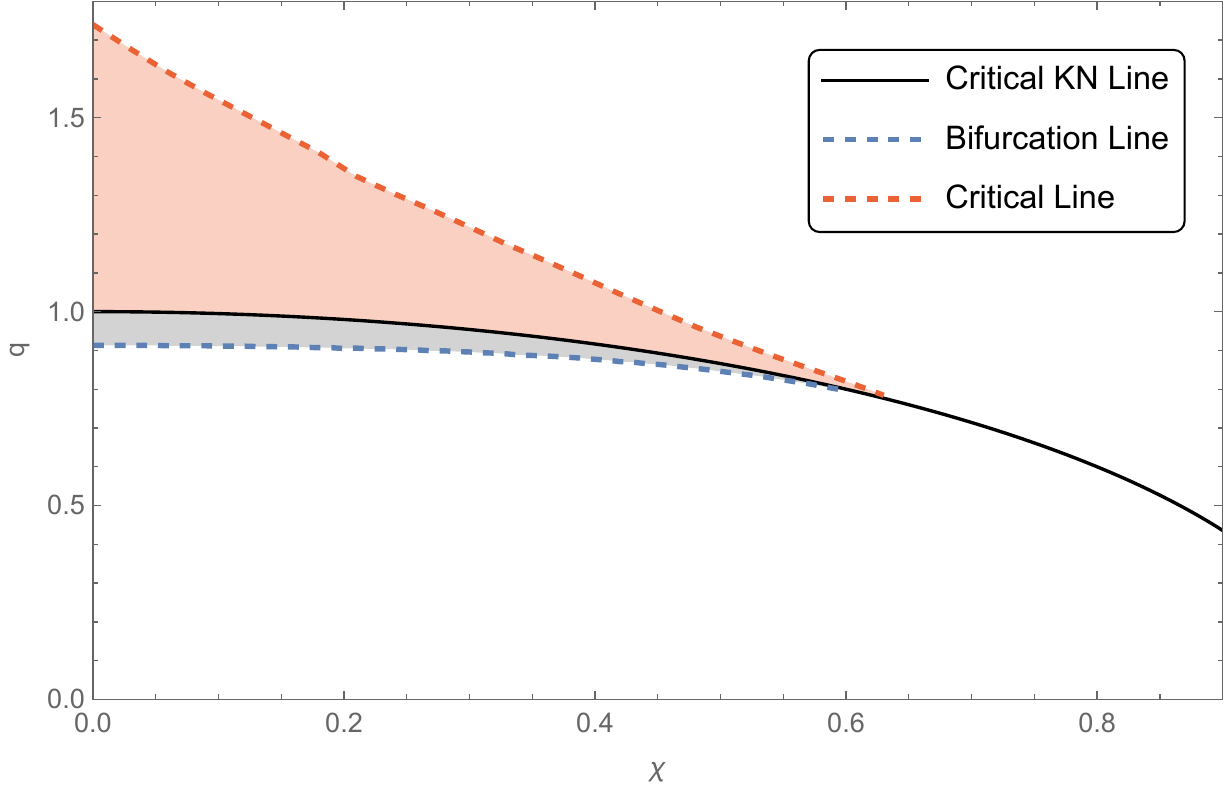}
    \caption{Domain of existence for the fundamental scalarization BHs with $b=-20$. The black line is the critical KN line below which the KN solutions exist. The blue dashed line is the bifurcation line where the cold and hot BHs branch from each other. The red dashed line depicts the critical line for the hot branch. The red region represents the domain where only hot black holes exist and the purple region demonstrate the co-exist area of the three branches.}
    \label{fig:domain of existence}
\end{figure}

We present the domain of existence of the scalarized BH solutions, in a charge $q$ $vs.$ spin $\chi$ diagram, for the quartic coupled function $e^{-b\phi^{4}}$ with $b=-20$ in Fig.\ref{fig:domain of existence}. We herein focus on the fundamental scalarized BHs without nodes. The nodes denote the zero points of the scalar field for the scalarized BH. All of the excited solutions with nodes are found to be unstable under radial perturbations and hence neglected\cite{Blazquez-Salcedo:2020nhs}. The black solid represents a critical line below which the KN BHs exist. There are two differently colored regions above and below the black line, which arises from the specific solution structure with respect to the quartic coupling. Jose $et\ al.$ has found that this model in the static limit $\chi=0$ has three distinct families of BH solutions, labeled as the bald BHs (RN BHs), the hot scalarized BHs, and the cold scalarized BHs\cite{Blazquez-Salcedo:2020nhs}. The red region marked in Fig.\ref{fig:domain of existence} indicates the area where only the hot BHs exist. The red dashed line (critical line) denotes the upper boundary of the red region and the hot BHs become extremal as it approaches. The blues dashed line (bifurcation line) together with the black line bound the gray region wherein the three families of BHs co-exist.

\begin{figure}[htbp]
    \centering
    \includegraphics[width = 0.4\textwidth]{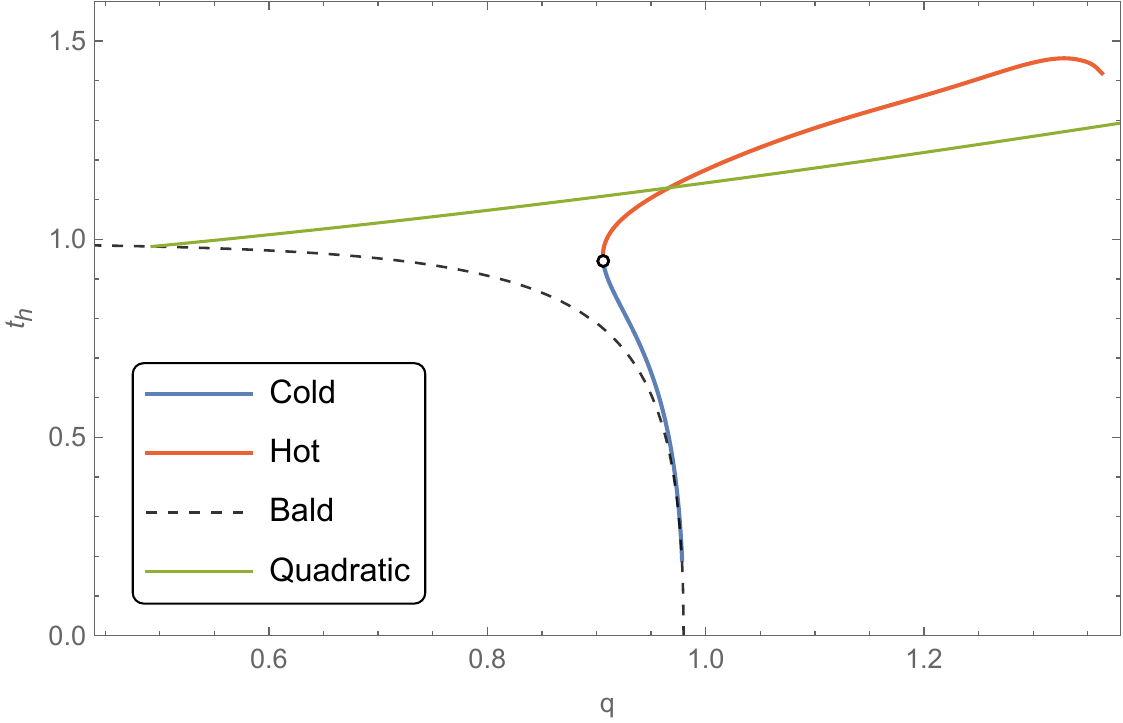}
    \includegraphics[width = 0.4\textwidth]{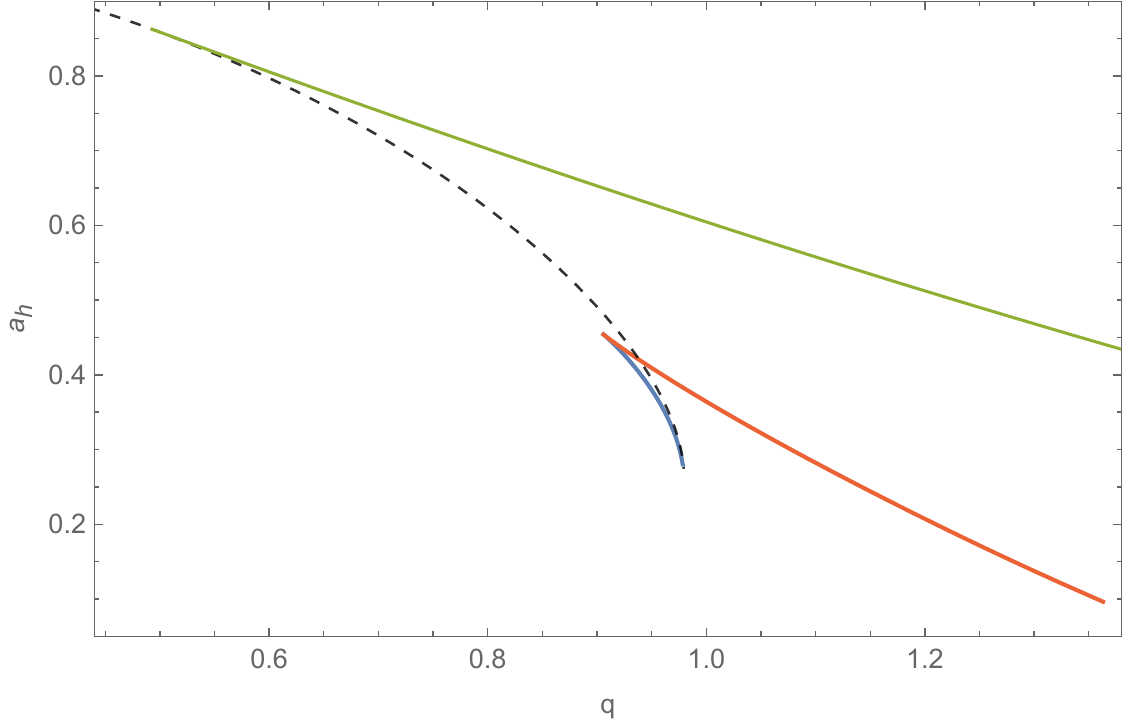}
    \includegraphics[width = 0.4\textwidth]{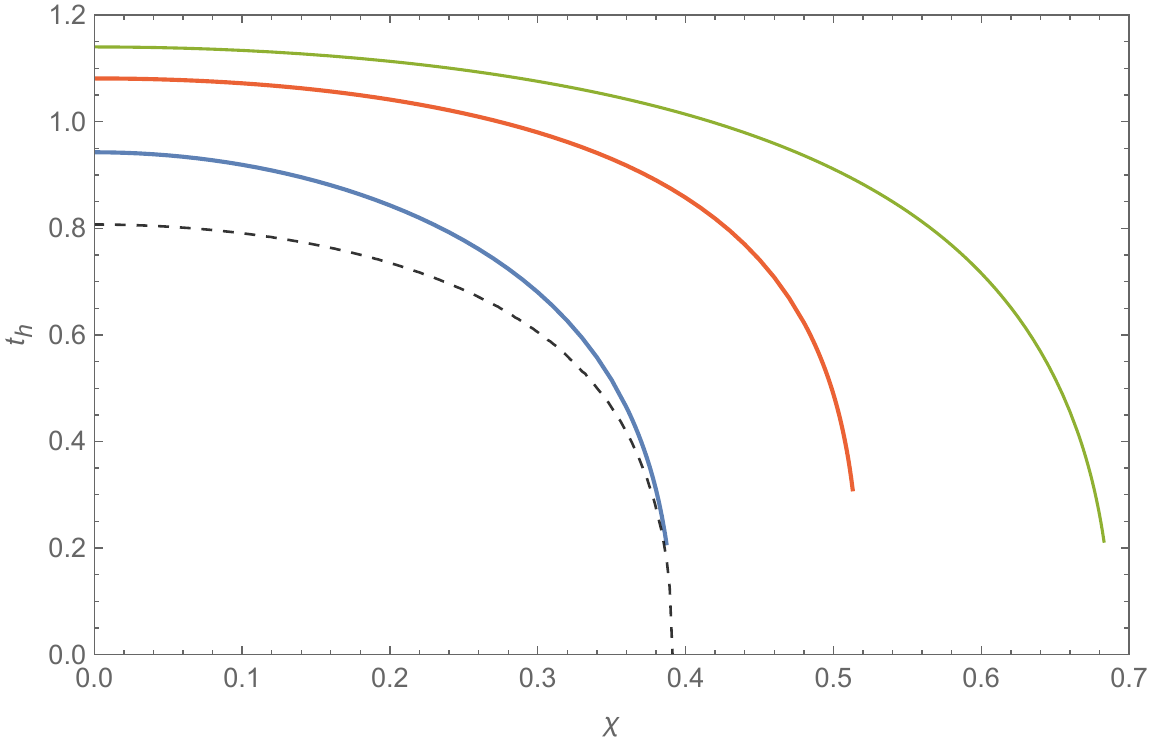}
    \includegraphics[width = 0.4\textwidth]{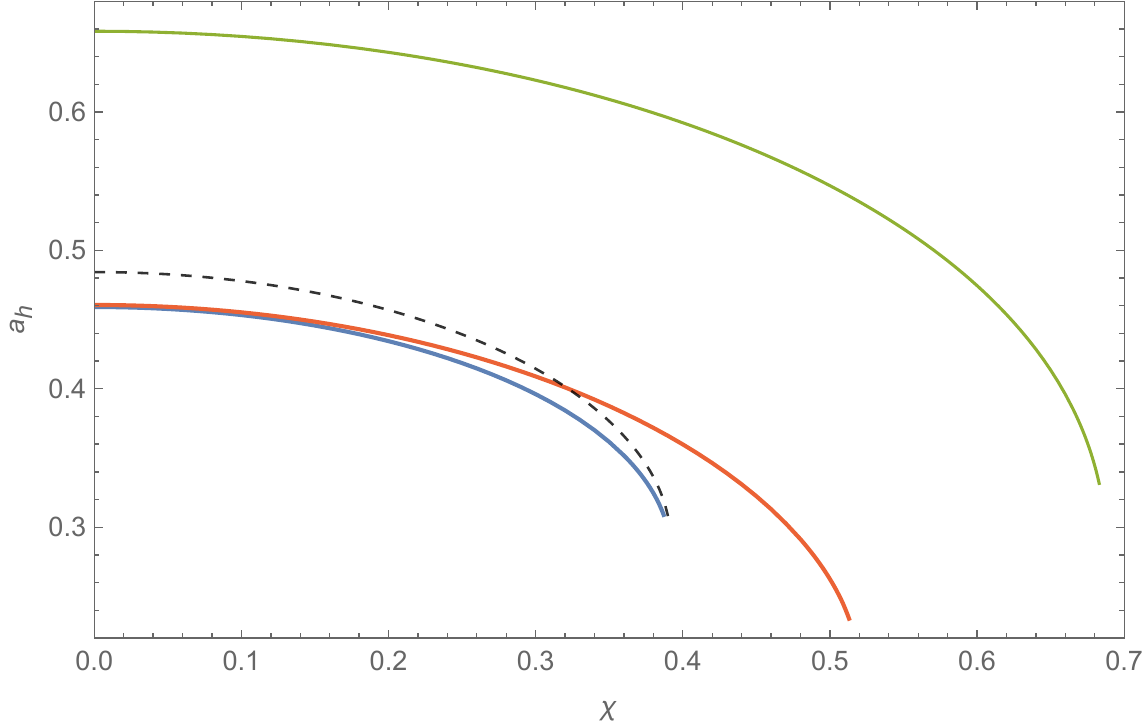}
    \caption{The thermodynamic behaviors decribed by the reduced temperature $t_{h}$ (left) and the reduced horizon area $a_{h}$ (right) of the three branches while varying $q$ with fixed $\chi=0.2,b=-20$ (upper) or varying $\chi$ with fixed $q=0.92,b=-20$ (bottom). The three branches of BH in the quartic EMS model is depicted by blue (cold), red (hot), black dashed (bald) respectively. The green line represents the hairy BHs in the quadratic EMS model for comparison. }
    \label{fig:thermodynamics}
\end{figure}

We demonstrate a complementary investigation in Fig.\ref{fig:thermodynamics} from a thermodynamic perspective. The cold, hot, and bald BHs are uniformly represented as the blue line, red line, and black dashed line, respectively. We present the results of the quadratic coupled EMS model ($f(\phi)=e^{-b\phi^{2}}$) as demonstrated by the green line in all four panels for the purpose of comparison. In the two upper panels of Fig.\ref{fig:thermodynamics}, we illustrate the behavior of the reduced temperature $t_{h}$ (left) and the reduced horizon area $a_{h}$ (right) for the three families of black holes while varying $q$ and fixing $\chi=0.2$, respectively. We find that the rotating scalarized BHs exhibit similar behavior to what they have in the static limit\cite{Blazquez-Salcedo:2020nhs}. The reduced temperature of the hot branch, starting from the bifurcation point, gets larger with increasing $q$ and undergoes a slight decrease toward the extremal $q$. For the cold branch, $t_{h}$ displays a monotonous increase as $q$ decreases, and connects to the hot branch at the bifurcation point smoothly. The cold branch seems to connect with the critical KN BH. However, as demonstrated in \cite{Blazquez-Salcedo:2020crd}, the static cold BHs become an intriguing critical solution with a nontrivial scalar hair while $q \rightarrow 1$. This critical solution is expected tending to possess a zero reduced temperature, similar to the case in the static limit. However, due to the numerical limitation, we can not illustrate the sharp drop up to zero for $t_{h}$ of the cold BHs as $q$ approaches the extremal value.  As implied by their name, the reduced temperature of hot BHs is always higher than that of cold BHs at the same parameters. 

The reduced horizon area $a_{h}$ of both the cold and hot branch exhibits the monotonous decrease in the right upper panel of Fig.\ref{fig:thermodynamics} while $q$ increases. $a_{h}$ of hot BHs is always larger than $a_{h}$ of cold BHs, indicating that hot BHs are more thermodynamically stable than cold BHs. On the other hand, $a_{h}$ of the hot branch is smaller than that of bald BHs while q is not large enough, and surpasses $a_{h}$ of bald BHs as $q$ increases. It suggests that hot BHs are thermodynamically unstable with small $q$ and become thermodynamically preferable while $q$ is sufficiently large. However, thermodynamic instability does not necessarily imply dynamical instability. It has been confirmed that the only branch among the three families with linear instability is the cold branch\cite{LuisBlazquez-Salcedo:2020rqp}. The hot and bald branch is stable under perturbations. In the fully interacting, spherically symmetric, nonperturbative, nonlinear dynamical evolution, the bald BH (RN BH) can evolve into a hot BH under sufficiently large perturbation\cite{Zhang:2021nnn}. There is a parameter controlling the initial perturbation strength associated with the threshold of this transition. The evolving solutions are attracted to a cold BH (also referred to as critical solution) and preserve for a certain time scale while the parameter for controlling the initial perturbation strength approaches sufficient to the threshold\cite{Jiang:2023yyn}. The similar thermodynamic behavior of axisymmetric rotating cold and hot BHs suggests that they possess similar dynamical properties. 

In summary, cold BHs emerge at the critical KN line, vanishing at the bifurcation line. Hot BHs branch from the bifurcation line with cold BHs and exist throughout the entire scalarized region in Fig.\ref{fig:domain of existence}. In the quadratic coupling model, scalarized BHs emerge at the bifurcation line from bald black holes, rather than at the KN critical line\cite{Guo:2023mda}. By adjusting the coupling constant, one can change the bifurcation line in the quadratic coupling model. The difference in the connection between hairy and bald solutions determines the dynamical dissimilarities between the quartic and quadratic coupling models.

The behavior of $t_{h}$ and $a_{h}$, while varying spin $\chi$ and keeping $q=0.92$ fixed, is illustrated in the bottom two panels of Fig.\ref{fig:thermodynamics}. $t_{h}$ of both cold branch and hot branch declines monotonically with increasing $\chi$. The cold branch converges to the extremal KN BH while $\chi$ approaches its maximum value for KN BHs. $a_{h}$ of the two families branch share the similar behavior of $t_{h}$ line. 
The $\chi$ enhances the difference of $t_{h}$ or $a_{h}$ between the cold and hot branch, i.e., $\chi$ amplifies the distinction between the two branches. However, the addition of spin does not fundamentally alter the thermodynamic comparison between cold and hot BHs. 

\begin{figure}[htbp]
    \centering
    \includegraphics[width = 0.4\textwidth]{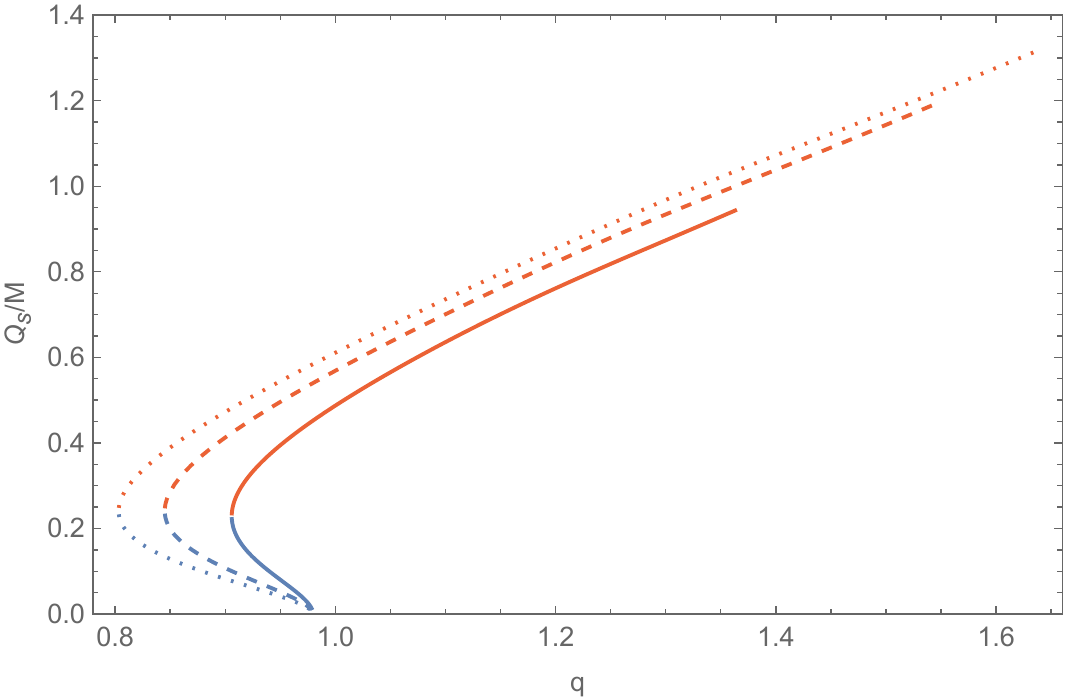}
    \includegraphics[width = 0.4\textwidth]{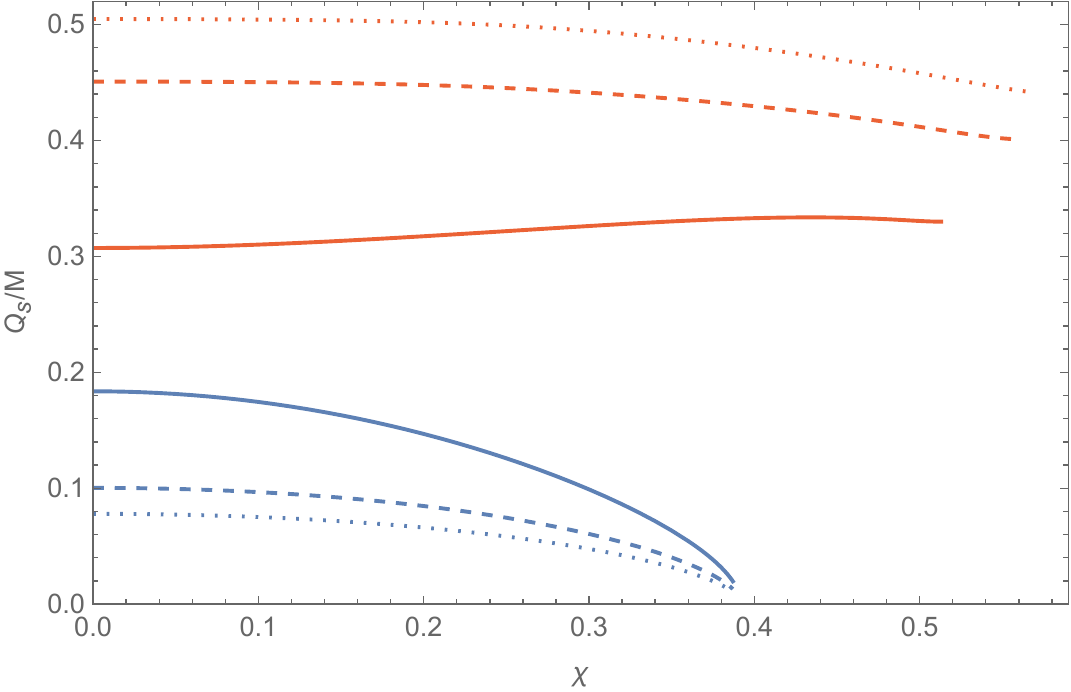}
    \caption{The scalar charge of the cold (blue) and hot (red) branches with varying $q$ and fixed $\chi=0.2$ in the left panel or varying $\chi$ and fixed $q=0.92$ in the right panel. The solid, dashed and dotted line show the behavior of $Q_{s}$ with different coupling parameter $b=-20,-40,-60$ respectively.}
    \label{fig:Qs}
\end{figure}

We also demonstrate the scalar charge of the cold (blue) and hot (red) branches with different $q$ or $\chi$ in Fig.\ref{fig:Qs} and depict the results with $b=-20\ \textrm{(solid)},\ -40\ \textrm{(dashed)},\ -60\ \textrm{(dotted)}$ respectively. The scalar charge of hot BHs increase monotonically with increasing $q$, while the scalar charge of cold BHs emerges from the bald BH and smoothly continues to hot BHs at the bifurcation point. We found that spin $\chi$ can suppress the scalar charge of cold BHs, but its influence on $Q_{s}$ of hot BHs is reletively small. It is also observed that a small coupling constant $b$ expands the existence range of scalarized solutions, but it affects the cold and hot branches differently. The decreasing $b$ increases the scalar charge of hot branch while reduce it for cold branch.

\begin{figure}[htbp]
    \centering
    \includegraphics[width = 0.4\textwidth]{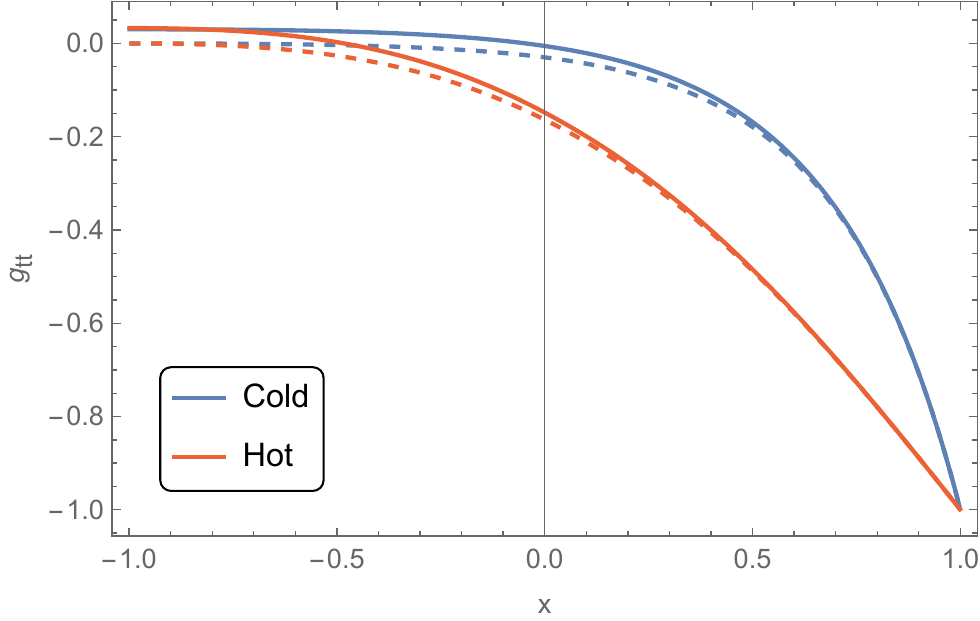}
    \includegraphics[width = 0.4\textwidth]{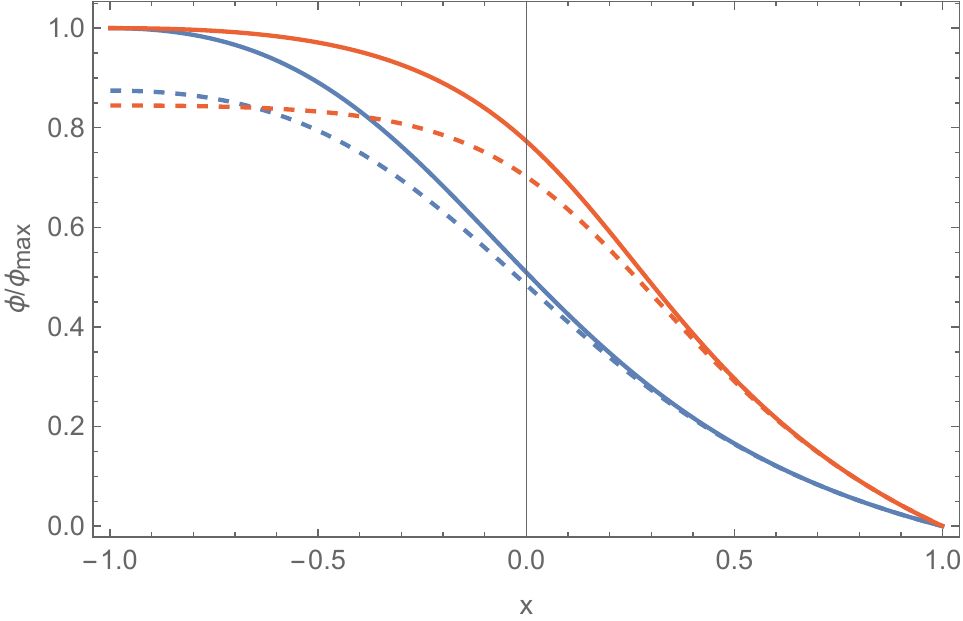}
    \caption{We depict the radial profile of the metric component $g_{tt}$ (left) and the scalar field $\phi$ (right) where $\chi=0.2,q=0.97,b=-20$. The blue (red) line represents the cold (hot) BH and the solid (dashed) line denotes the profile with the fixed angular coordinate $\theta=\pi/2$ ($\theta=0$). The $g_{tt}$ values at the same angles for different branches may seem identical at the horizon ($x=-1$) in the left panel. However, there are subtle differences between them.}
    \label{fig:profile}
\end{figure}

We show the radial profile of $g_{tt}$ metric component and scalar field for the cold (blue) and hot BH (red) with $\chi=0.2,q=0.97$ in Fig.\ref{fig:profile}. The solid line and dashed line represent the radial profile at $\theta=\pi/2$ and $\theta=0$ respectively. The value of scalar field is divided by its maximum, which is always the value at location $(x=-1,\theta=\pi/2)$, within the $(x,\theta)$ domain. The scalar hair of the hot BH is more widely distributed outside the horizon compared to the cold BH. The scalar profile of the cold branch, on the other hand, is concentrated near the horizon. 

The difference between the solid and dashed lines demonstrates that the scalar field is more accumulated on the equatorial plane $\theta=\pi/2$ than the rotation axis $\theta=0$. 

\begin{figure}[htbp]
    \centering
    \includegraphics[width = 0.5\textwidth]{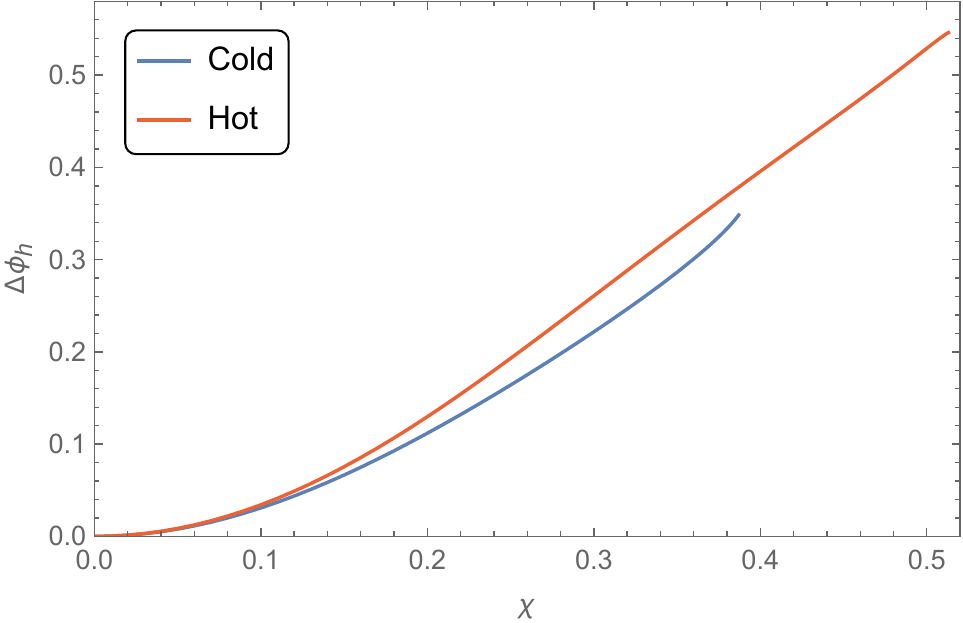}
    \caption{The relative difference defined as $\Delta \phi_{h} \equiv (\phi(-1,\pi/2)-\phi(-1,0))/\phi(-1,\pi/2)$ with $q=0.92,b=-20$.}
    \label{fig:dphi}
\end{figure}

\begin{figure}[htbp]
    \centering
    \includegraphics[width = 0.4\textwidth]{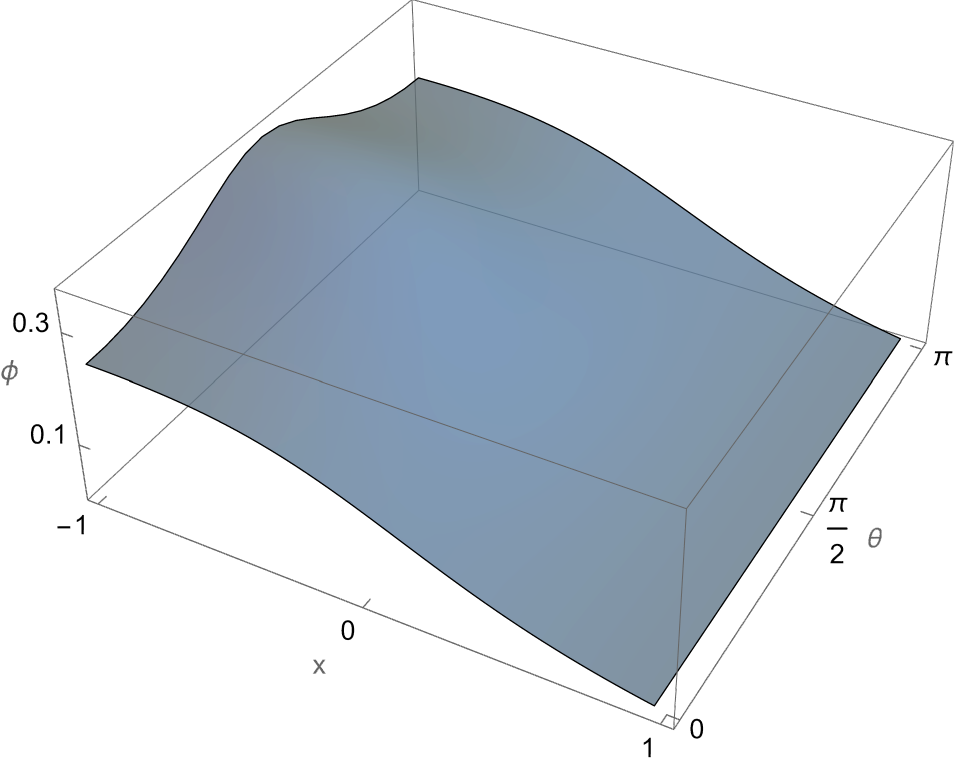}
    \includegraphics[width = 0.4\textwidth]{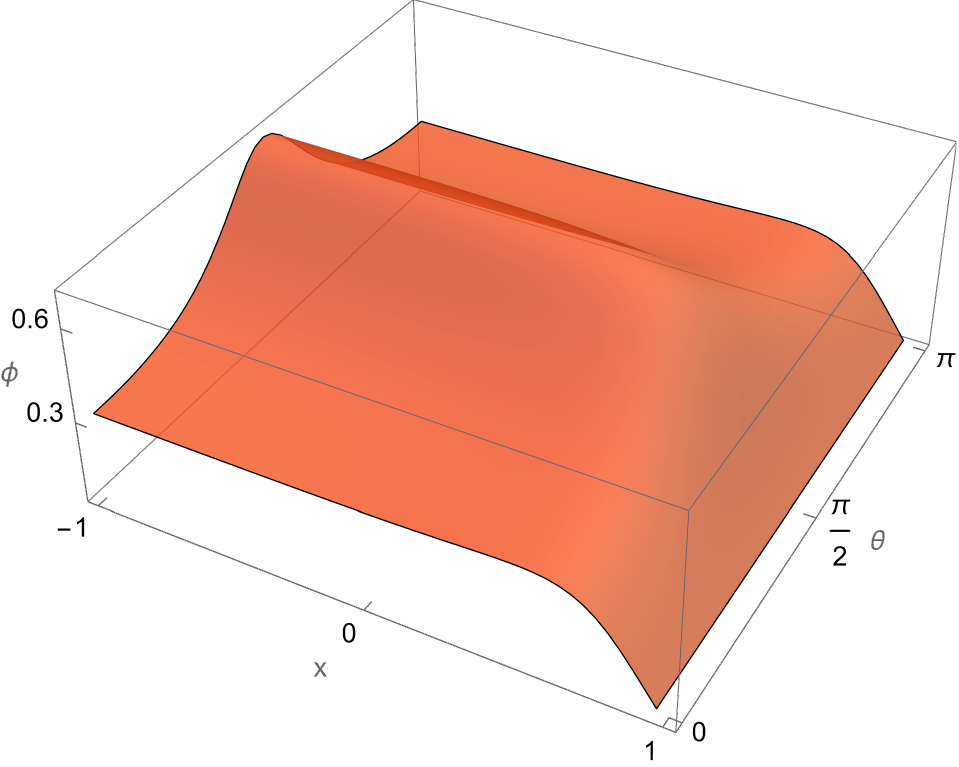}
    \caption{The 3d profile of scalar field for cold (left) and hot (right) branch on the $(x,\theta)$ domain. The parameters of the cold BH is given by $\chi=0.387,q=0.92,b=-20$ and the hot BH is fixed $\chi=0.513,q=0.92,b=-20$. }
    \label{fig:3d}
\end{figure}

We define the relative difference $\Delta \phi_{h} \equiv (\phi(-1,\pi/2)-\phi(-1,0))/\phi(-1,\pi/2)$ on the horizon to quantify the variation for scalar condensation at different elevation. The relative differences with different $\chi$ for the cold and hot branches are shown in Fig.\ref{fig:dphi}. $\Delta \phi_{h}$ monotonically increases with increasing $\chi$ for both branches as expected. The scalar profile of the cold BH (left) and the hot BH (right) on the entire domain is presented in Fig.\ref{fig:3d} for an intuitive perspective, which $\chi$ is chosen to be near the extremal spin value for each branch. The scalar field peak of the hot branch on the equatorial plane is sharper near extremality compared to the cold branch. 

\begin{figure}[htbp]
    \centering
    \includegraphics[width = 0.5\textwidth]{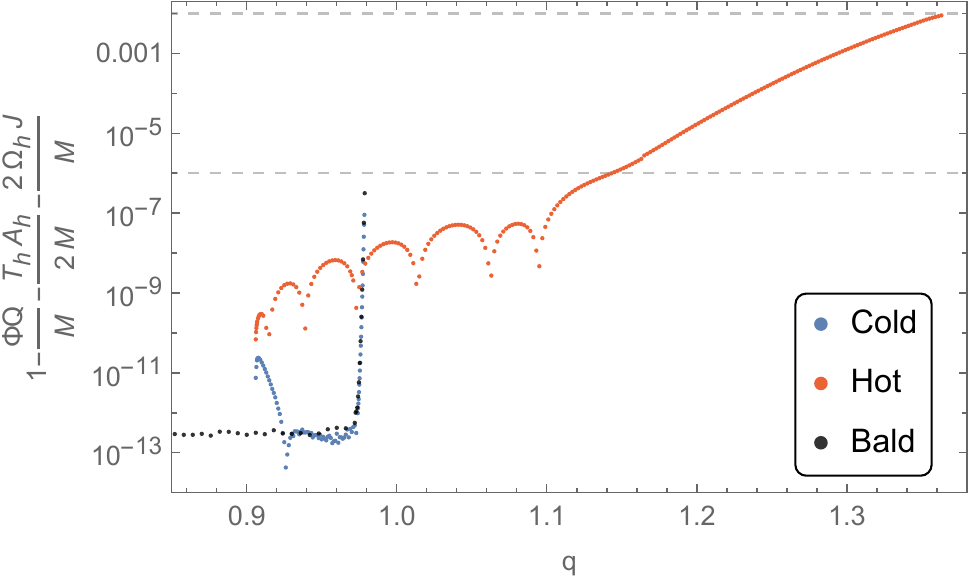}
    \caption{The error estimation described by the Smarr relation (\ref{eq:Smarr}) with different $q$ and fixed $\chi=0.2,b=-20$. The error of cold, hot and bald BHs are shown by the blue, red and black dots individually. The two gray dashed line represent $10^{-6}$ line and $10^{-2}$ line respectively, which denotes the maximum error for different branches.}
    \label{fig:error}
\end{figure}

We demonstrate the Smarr relation as an error estimation for each of the three branches in Fig.\ref{fig:error}, with $\chi=0.2$ fixed and varying $q$. The errors of cold and bald BHs are mostly limited to less than $10^{-10}$, rapidly rising to $10^{-6}$ when $\chi$ approaches the extremal value. The errors for hot BHs are larger than other branches, especially for the overcharged hot BHs. However, in general, the errors manifested by the Smarr relation do not exceed $10^{-2}$.

\section{Discussion}
\label{section4}

In this paper, we investigated the rotating scalarized BH solutions in the EMS model with quartic coupled function $e^{-b\phi^{4}}$. We regained three branches (cold, hot, and bald) of BH solutions which have been identified in the static limit, and presented the domain of existence for the cold and hot scalarized branches on the $(\chi,q)$ plane with $b=-20$. The thermodynamic behavior of rotating scalarized BHs ($\chi=0.2$) was demonstrated with different $q$, which is similar to the scalarized BHs in the static limit as expected. The effect of spin $\chi$ on the thermodynamic behavior of each branch was depicted with fixed $q=0.92$, where $\chi$ enhances the difference between the two scalarized branches. We also showed the radial profile of the metric component $g_{tt}$ and the scalar field $\phi$ for the two branches to offer an intuitive perspective. The scalar field of the hot branch is more concentrated in the region outside the event horizon, indicating that it occupies a larger portion of the system energy compared to the cold branch. The increasing spin amplifies the differences of the scalar field between cold and hot branches at different angular, as demonstrated in Fig.\ref{fig:dphi}. The 3D profiles of the scalar field, while $\chi$ approaches its extremal value with $q=0.92$ for each branch respectively, were presented in Fig.\ref{fig:3d}. We also manifested the error estimation established by the Smarr relation with increasing $\chi$. The Smarr relation of bald (KN BHs) and cold branches do not exceed $10^{-6}$ even when approaching the critical KN line. However, the errors for hot BHs become relatively large while the hot BHs overcharged, yet they remain below $10^{-2}$.

While this paper discusses the unique thermodynamic behavior of scalarized BHs in the quartic coupling EMS model, it must be pointed out that the linear stability analysis of scalarized BHs for different branches should be validated using perturbation theory. What is more, the nonlinear instability of KN BHs beyond spontaneous scalarization in this model, as well as the critical phenomenon, require the implementation of dynamic evolution techniques on the background of axisymmetric stationary spacetimes. This is indeed extremely complex, but it is not impossible because of the inherent simplicity of the EMS model itself. Besides the BH entropy comparison between the hot and bald branch (Fig.\ref{fig:thermodynamics}) suggests that the thermodynamically unstable does not exclusively point to the dynamic instability. We observed that the bald branch and hot branch do not connect, instead bridged by the cold branch. The key point seems to be whether the two compared branches are directly connected. The cold BHs have lower entropy compared to its two smoothly connected branches, and hence are dynamic unstable. To clarify this question further, a complete investigation is required, which is beyond the scope of this paper. 

Let us put attention back to EMS model itself. The spin-induced instability of KN BHs with quadratic coupling has been identified\cite{Hod:2022txa}, revealing the existence of spin-induced spontaneous scalarization in EMS model\cite{Lai:2022ppn,Lai:2022spn}. The spin-induced scalarized BH solutions have yet to be calculated, which are expected to feature the arbitrarily small (though not zero) electric charge and nonzero spin, even with a significantly large coupling constant. Investigations on such hairy solutions can contribute to constraint EMS model using current astronomical observations.

\begin{acknowledgments}
	The work is in part supported by NSFC Grant No.12205104, No.12375048, ``the Fundamental Research Funds for the Central Universities'' with Grant No.  2023ZYGXZR079, the Guangzhou Science and Technology Project with Grant No. 2023A04J0651 and the startup funding of South China University of Technology.
\end{acknowledgments}

\end{document}